\documentclass[conference]{IEEEtran}

\usepackage{balance}  
\usepackage{graphicx} 
\usepackage{times}    
\usepackage{url}      
\usepackage[utf8]{inputenc}
\usepackage[usenames, dvipsnames]{color}
\usepackage{balance}
\usepackage{moreverb}
\usepackage{amsmath}
\usepackage[utf8]{inputenc}
\usepackage{url}

\usepackage{listings}
\usepackage{color}
\usepackage{textcomp}
\usepackage{algpseudocode}
\usepackage{algorithm}
\usepackage{lmodern}
\usepackage{enumerate}
\usepackage{varwidth}
\usepackage{xcolor}
\usepackage{xspace}

\definecolor{mygreen0}{rgb}{0, 0.75, 0}

\definecolor{myred1}{rgb}{1,0,0}
\definecolor{mygreen1}{rgb}{0, 1, 0}
\definecolor{myblue0}{rgb}{0, 0, 1}

\definecolor{myred2}{rgb}{1,0.5,0.5}
\definecolor{mygreen2}{rgb}{0.5, 1, 0.5}
\definecolor{myblue2}{rgb}{0.5, 0.5, 1}

\definecolor{mygreen}{rgb}{0, 0.25, 0}
\definecolor{myblue}{rgb}{0, 0, 0.75}
\definecolor{myred0}{rgb}{0.5,0,0}

\definecolor{listinggray}{gray}{0.98}
\definecolor{lbcolor}{rgb}{0.98,0.98,0.98}
\newcommand{\projectName}{\emph{SMART}\xspace}

\begin{document}
\date{}
\bibliographystyle{IEEEtran}
\title{\vspace{-0.5em}\LARGE \bf Not Every Flow is Equal -- \\SMART Discrimination in Redundancy\vspace{-1.5em}}

\author{{\bf Pradeeban Kathiravelu} \\
{\small INESC-ID Lisboa / Instituto Superior Técnico} \\
{\small Universidade de Lisboa, Portugal}\\
{\small pradeeban.kathiravelu@tecnico.ulisboa.pt\vspace{-1.5em}}\\
\and 
{\bf Lu{\'\i}s Veiga} \\
{\small INESC-ID Lisboa / Instituto Superior Técnico} \\
{\small Universidade de Lisboa, Portugal}\\
{\small luis.veiga@inesc-id.pt	\vspace{-1.5em}}}
\maketitle

\subsection*{Abstract}
Software-Defined Data Centers (SDDC) extend virtualization, software-defined networking and systems, and middleboxes to provide a better quality of service (QoS). While many network flow routing algorithms exist, most of them fail to adapt to the dynamic nature of the data center and cloud networks and their users' and enterprise requirements. This paper presents SMART, a Software-Defined Networking (SDN) middlebox architecture for reliable transfers. As an architectural enhancement for network flows allocation, routing, and control, SMART ensures timely delivery of flows by diverting them to a less congested path dynamically in the software-defined data center networks. SMART also clones packets of higher priority flows to route in an alternative path, along with the original flow. Hence SMART offers a differentiated QoS through varying levels of redundancy in the flows.
\vspace{-0.7em}

\section{Introduction}
\label{sec:intro}
Enterprise data centers are designed to offer high-availability and fault-tolerance, abiding to service level agreements (SLA). Efficient and high performance network topologies~\cite{shin2011small} are often tailored for the specific characteristics and requirements of the data center. SDN offers flexibility and configurability to data center networks~\cite{sezer2013we}, while middleboxes manage the load balancing, policy control, and security aspects of the data center~\cite{joseph2008policy}.

While ensuring lower monetary cost and improving energy and carbon-efficiency (ECE)~\cite{khosravi2013energy}, data should be transferred abiding the SLA~\cite{simao2013flexible}. Data locality in the networks is further driven by the geopolitical and customer requirements. Network flows in data centers consist of flows of packets of different priorities and deadlines from multiple users~\cite{wilson2011better}. Priority flows often have stricter SLA deadlines to be met. Existing networks generally utilize routing algorithms that often do not consider any SLA, system policies, and user preferences. 

Software and hardware middleboxes provide specific custom functions and important features crucial to the network~\cite{gember2012toward}, and hence cannot be eliminated from the data center deployments. Research proposes efficient architectures to mitigate the potential overheads imposed by the middleboxes~\cite{walfish2004middleboxes}, such as the seamless middlebox deployments enabled by SDN~\cite{gember2012toward}, offering complimentary features to the network. Leveraging and extending recent middlebox and SDN research and developments, flows can be tagged with custom information, that can be read and interpreted by the applications deployed on top of the northbound API of the controller~\cite{fayazbakhsh2014enforcing}. Thus information on SLA, business rules, and policies can be included as custom headers with the packets.

Middleboxes can be part of an SDN, or deployed separately anywhere, with the centralized control offered by SDN~\cite{gember2012toward}. FlowTags proposes an extended SDN and middlebox architecture that offers dynamic functionality to the SDN, by adding custom tags to the packets, to enforce network-wide policies, providing flow tracking capabilities~\cite{fayazbakhsh2014enforcing}. Slick proposes a control plane for middleboxes, extending the SDN paradigm and architecture to network middleboxes~\cite{anwer2013slick}. Convergence of middleboxes and SDN has provided many advantages including flexibility in middlebox placement, effective failure handling, scalability~\cite{qazi2013practical}, and efficient policy enforcement~\cite{qazi2013simple}.

This paper presents SMART, an approach of adaptive redundancy in a set of flows, focusing to fulfill differentiated levels of SLA across the flows. Priority flows are tagged to indicate thresholds such as maximum routing time and other user-defined QoS parameters at the origin node by SMART. Tags will be read and interpreted at the intermediate nodes as policies, and controller will be triggered upon violation, adhering to the SDN paradigm. Thus, tags are used in detecting potential network congestion, SLA violation, or delays in routing the priority flows. When the controller is triggered for such violation, the packets from a subflow of the flow, possibly of various lengths is diverted in an alternative route to the destination, or the subflow is cloned and routed in an alternative route along with the original flow.

In the upcoming sections, we will further analyze the proposed SMART approach of differentiated redundancy. Section II discusses the design and solution architecture of SMART and elaborates the prototype implementation. Preliminary evaluations on SMART are discussed in Section III. Section IV briefly discusses the related work. Finally, Section V closes the paper discussing the current state of the research and future work. 
\vspace{-0.5em}

\section{\projectName Discrimination of Network Flows}
\label{sec:arch}

SMART is a software middlebox architecture that enforces a set of enhancements over existing network flow routing algorithms, leveraging the northbound API of an SDN controller and exploiting the functionality of adding tags to the packets proposed by FlowTags~\cite{fayazbakhsh2014enforcing}. FlowTags is extended to provide SLA-awareness to the flows with minimal overhead, as no other existing SDN-based approach enables per-flow custom policy enforcement in a network with presence of software and hardware middleboxes.

SMART exploits the monitoring capabilities offered by SDN, while extending the complimentary features offered by middleboxes to mark the priority flows. Packets of priority flows consist of the information on SLA parameters, in the form of tags attached to the packets extending and leveraging FlowTags to indicate parameters relevant for the SLA enforcement of the flow. Packets of the priority flows are tagged at the origin node by the distributed SMART software middlebox architecture consisting of FlowTagger deployed on the nodes, and the tags are read at the nodes en route destination, by the FlowTagger. Priority flows can be all the flows of a given user, all the flows originating at a given node, or a set of flows that meet a certain user-defined custom criteria.

SLA parameters contain hard limits or thresholds such as the maximum permissible routing time, and also soft limits that are fractions of the respective hard limits. Upon encountering a soft limit, based on the policy and the length and the priority level of the flow, either the flow is replicated and rerouted from its origin to the destination in an alternative route, or it is cloned or diverted from a break point node partially, to mitigate the potential SLA violation by certain malfunctioning or congested nodes in the initial route. With a little redundancy, SMART attempts to meet the deadlines of the priority flows.
			\vspace{-1em}

\subsection{Software Architecture and Design}
\begin{figure}[ht]
	\begin{center}
		\resizebox{\columnwidth}{!}{
			\includegraphics[width=\textwidth]{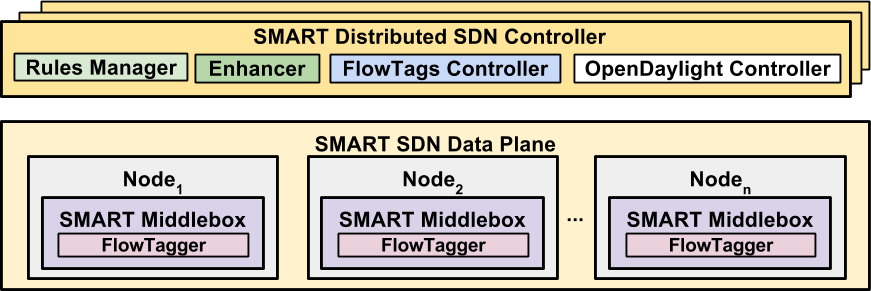}
		}
		\vspace{-2.5em}
	\end{center}
	\caption{SMART Deployment Architecture}
	\vspace{-0.5em}
	\label{fig:arch}
\end{figure}

Figure~\ref{fig:arch} depicts the higher level deployment architecture of SMART. OpenDaylight is extended and used as the base SDN controller, physically distributed across a cluster of computers, with a logically centralized view.

	\vspace{-1em}

FlowTags architecture has been extended and adapted as a software middlebox inside each node of the data plane as well as a FlowTags controller. The SMART middlebox consisting of the FlowTagger resides inside the nodes that are the origins of the flows. FlowTagger reads and writes tags to the packets. Tags include current time stamp to track the time consumed in routing so far which can be used to estimate other optional information such as estimated monetary cost and energy consumption. 

FlowTags controller is designed as an extension to OpenDaylight SDN controller to parse the tags from the packets forwarded to the controller, and also to control and invoke the FlowTagger to tag the packets of the flows originating from a node. Policies, thresholds, and business rules are read and stored into the SMART controller from the configuration files, as defined in the network by the system administrators or the users, and managed by the rules manager. Rules manager reads the rules from the tags, and triggers the SMART enhancer according to the defined policies. SMART enhancer consists of the enhancement algorithms on top of the base routing algorithms, as an extension to the controller. 

Along with the other rules set by the SDN controller, the custom user-defined tags in the packets read from FlowTagger are interpreted as policies, which the packets should respect. Upon the violation of the policies in any of the nodes, the first packet of the violating flow is sent to the SMART controller and triggers it. Then controller sets a break point in the flow on the packet that triggered the controller and the location node of the packet when the violation occurred. The break point node or packet can also be chosen algorithmically.

	\vspace{-1em}

\subsection{SMART Enhancement Mechanisms}
SMART offers 3 alternative approaches - divert, clone, and replicate, based on the priority level of the flow, to provide an SLA-aware data center network. Algorithmic improvements handle these cases based on the nature of the flow and the defined policies. Table~\ref{table:overhead} presents the potential overhead for the priority flows in completion time and bandwidth usage. Time overhead is measured as a potential delay that may happen in the flow delivery compared to the original flow. Here n refers to the number of times the packets are cloned, diverted, or replicated.

\begin{table}[!t]
\caption{Time and Bandwidth Overhead}
\label{table:overhead}
\begin{tabular}{|c||c| |c|}
\hline
Approach & Duplicate Packets & Time Overhead \\

\hline
Divert(n) & (n-1) * (0 -- 100)\% & Possible   \\
Clone(n) & n * (0 -- 100)\%  & No/Negligible\\
Replicate(n) & n * 100\% & No/Negligible\\
\hline
\end{tabular}
			\vspace{-2em}

\end{table}

\paragraph*{\textbf{Flow Diverting Approach}}
In the diverting approach, subflows of a few selected priority flows are routed in an alternative path to the destination, when an SLA violation is expected due to a congested node or link in the original route. The subflow is diverted in a single or multiple alternative routes except the original route. Choosing multiple alternative routes in divert approach will be useful for higher priority flows, when there is no certain alternative route that can be considered the best alternative. As the chosen routes may be longer or suboptimal than the original route, and as there is a need to reconstruct the flow, there is a potential time overhead or delay. Divert approach does not have duplicate packets if the subflow is diverted in just one alternative direction. However, diverting to multiple alternative directions will have duplicate packets, just like the cloning approach. 

\paragraph*{\textbf{Flow Cloning Approach}}
Clone approach clones the subflow to a single or multiple alternative routes. Cloning approach is employed for higher priority flows, where flows are cloned partially or fully, instead of merely diverting. The original flow is left to continue in its route, while subflows are cloned and routed in an alternative route towards the destination. The updated rule in the break point ensures sending the packets in the original route as well as an alternative route. As the original flow is left to continue in its original route unmodified, clone approach does not have a time overhead.

\paragraph*{\textbf{Flow Replicating Approach}}
In the cloning and diverting approaches, controller clones or diverts the packets that follow the break point packet respectively, by changing the routing rules for the packets of the priority flows in the break point node. Alternatively, the entire violating flow can be replicated from the origin to destination in a single or multiple alternative routes. Flow cloning and replicating are further enhancements to flow diverting, as the original route could be a better choice, if the congested links or nodes recovered during the transmission. Similar to the cloning approach, replicate approach do not have a time overhead as well. Replicating entire flows imposes 100\% of duplicate packets till the routing is complete. Replicate approach can also be extended to drop the original flow, as in the divert approach. While reducing the duplicate packets, this may introduce a time overhead.

	\vspace{-1em}

\subsection{SMART Enhancer Algorithms}
$SmartRoute$, the core routing procedure is described in Algorithm~\ref{alg:smartFlow}. The algorithm diverts or clones sub sets of priority flows, known as subflows, when the current routing fails to complete the transmission of the flow within the stipulated soft limit. These limits, set by the controller on the switches will trigger a communication to the controller from the switches when a violation is imminent. Soft limit parameters are often modeled as a fraction of the respective hard limit parameters, such as routing time. Tags such as priority and SLA parameters are added to the packets of the flows to provide the addition information required in accomplishing this.

			\vspace{-1.0em}

\begin{algorithm}[ht]
			\fontsize{9}{9}\selectfont			
	\caption{SMART Enhancement}
	\label{alg:smartFlow}
	\begin{algorithmic}[1]
\Procedure{SmartRoute}{$flow$, $origin$, $destination$}
		\Repeat
		\State \colorbox{green!10}{$BaseRoutingAlgorithm(flow, origin, destination)$}
		\State \colorbox{green!10}{$flow.status.update()$}
		\If {\colorbox{red!10}{($flow.policies.isThresholdMet()$)}}
		\State \colorbox{green!10}{$cloneOrigin$ $\gets$ $markBreakPoint$($flow$, $origin$,}\par\hspace{160pt}  \colorbox{green!10}{$destination$)} 
		\State \colorbox{green!10}{$cloneDestination$ $\gets$ $findCloneDestination($}\par\hspace{120pt}\colorbox{green!10}{$flow$, $flow.status$)}
		\State \colorbox{green!10}{\textit{clonedFlow} $\gets$ $cloneFlow(flow$, $cloneOrigin$,} \par\hspace{140pt}\colorbox{green!10}{cloneDestination)}
		\EndIf
		\Until{\colorbox{red!10}{($flow.allReceived(cloneDestination)$ \textbf{or}}} \par\hspace{80pt}\colorbox{red!10}{\textit{flow.allReceived(destination))}}
				\State \colorbox{green!10}{$mergeFlows$($flow$,$clonedFlow$)}
 		\State \colorbox{green!10}{\textit{dropPacketsOnTransmission(flow.parentID)}}	
		\EndProcedure
	\end{algorithmic}

\end{algorithm}
		\vspace{-1em}

The $BaseRoutingAlgorithm$ refers to any underlying routing algorithm such as Dijkstra's shortest path algorithm~\cite{dijkstra1959note} or equal-cost multi-path (ECMP) algorithm~\cite{hopps2000analysis}, which is to be enhanced by SMART. The thresholds can be defined as system-wide policies, such as minimal throughput and latency, in network system and individual flow level.

Statistics when routing through each link is monitored to offer fault-tolerance to the data center network. Nodes or links that take much longer time to route the flows or packets than the average time to route, those consume unconventionally large amount of energy or computing resources in routing, or those who exhibit a similar behavior that may lead to exceeding the threshold specified in the SLA, are considered to be functioning poorly, and acted upon. The status of the flow is updated to the controller as the tags are read by the middlebox architecture. 

SmartRoute routes the flows from origin to destination entirely using BaseRoutingAlgorithm, unless the threshold is met. The SMART software middlebox is triggered to reroute the subflow in the new alternative route towards the destination, or is invoked to forward the packets of the subflow to both the original and alternative route by the SDN controller, when an SLA violation for the flow is imminent according to the policies, or when the threshold defined in the flow policies is met.

A node and a packet are chosen as the break point node and packet respectively. Having the break point node as the origin, a subflow is cloned or diverted starting from the break point packet to the rest of the flow. The destination of the cloned or diverted subflow is defined as the clone destination, where the subflow is merged with the rest of the flow to reconstruct the original flow.

\paragraph*{\textbf{Clone Destination}} 
Clone destination is either the destination of the original flow, or the next node following the congestion. The procedure findCloneDestination() decides the clone destination based on the flow and its status, which consists of further information which can be used to find the nature of the policy violation. When a congestion is encountered, the exact destination is decided based on the characteristics of the congestion. In a large data center with a few nodes identified to be contributing to the congestion, the cloned or diverted subflow is routed towards the node that immediately follows the congested link to avoid routing in a sub-optimal path when the congestion affects just one or a few of the nodes in the original route. This also minimizes redundant packets by enabling early recomposing of the original flow. If there is no such nodes identified to be contributing to the congestion, the cloned or diverted subflow is routed towards the original destination in an alternative route. 


\paragraph*{\textbf{Flow Reconstruction}}

Once the entire packets of the flow, regardless whether from original or cloned flows, are received at clone destination, the original flow is reconstructed. If the clone destination is different from the original destination, the recomposed flow is left to continue in its original route towards the destination. Leveraging the FlowTags custom tags, duplicate packets are detected and dropped in the clone destination upon receiving the entire flow, ensuring end-to-end transmission guarantee. 

Cloning approach minimizes the necessity to reconstruct the flow, if the entire packets from the original flow are received before the packets from the clone, hence dropping the clone. For the diverting approach, and for the cloning approach if the packets of the cloned flows arrived earlier, the flow will be reconstructed by merging the packets from the cloned or diverted subflow to the packets of the original flow that have already arrived. 

The following priority flows of the same path may be replicated and rerouted, or diverted in the origin, in an alternative route. Thus, while the initial flows that are identified to violate the SLAs may still violate SLAs due to the time taken in cloning the flow, following flows will be able to avoid the violating route altogether. A replicate approach resends the entire flow from the origin to the destination in one or more alternative routes, which avoids the necessity for recomposing and packet-level manipulation, with more redundancy.

\paragraph*{\textbf{Break Point}}
Break point is a pointer to the node and flow where the subflow is cloned. The controller chooses the break point programmatically, and writes rules on the break point nodes to divert or clone the upcoming packets of the priority flows. While break point is crucial in the SMART enhancement algorithm, it is used just for the subflow construction, and information on break points are not stored statically in the flows or the controller beyond the time frame of subflow construction. 

Algorithm~\ref{alg:mark} elaborates marking a break point for the flow, which first needs deciding the exact node to be the break point node, and also find the exact packet from which the flow is to be included in the diverted or cloned subflow. If a specific node or link is estimated to be responsible for the policy violation, the node will be marked as the breakpoint node. If there is no such specific or explicit malfunctioning link or node to be blamed, the delay may be due to other factors such as network congestion across multiple nodes and links or the flows being much larger than the average flows in the data center and hence taking longer than expected. Here, the break points depend on policies or are decided statistically.

	\vspace{-1em}

\begin{algorithm} [ht]
				\fontsize{9}{9}\selectfont			
				
	\caption{Marking the Break Point}
	\label{alg:mark}
	\begin{algorithmic}[1]
		\Procedure{markBreakPoint}{$flow$, $origin$,\par\hspace{100pt} $destination$,$policies$, $links$}
		\For {	\colorbox{red!10}{($link$ \textbf{in} $flow.route$)}}
		\If 	{	\colorbox{red!10}{\textit{(policies.isThresholdMet(link.param))}}}\\
		\Comment{A clearly visible malfunctioning link exists}
		\State 	 \colorbox{green!10}{breakPoint.node $\gets$ current.node}
		\State 	 \colorbox{green!10}{breakPoint.packet $\gets$ current.packet}
		\State 	 \colorbox{blue!10}{\textit{\textbf{Return} $breakPoint$}}
		\EndIf
		\EndFor
		\State 	 \colorbox{green!10}{breakPoint $\gets$ flow.estimate(policies.breakPolicy)}		
		\State \colorbox{blue!10}{\textit{\textbf{Return} $breakPoint$}}
		\EndProcedure
	\end{algorithmic}
	\vspace{-0.3em}

\end{algorithm}
	\vspace{-1em}
As the origin of the diverted or cloned subflow, break point node reroutes the packets to the destination in an alternative route as they arrive at the node. All the following packets arriving to the break point node will be diverted in the alternative route, while the packets of the original flow following the break point is left to continue in the clone approach. 
\vspace{-1.8em}


\subsection{Prototype Implementation}
\label{sec:Implementation}
			\vspace{-0.7em}

A prototype of the proposed solution is implemented leveraging the OpenDaylight controller, while exploiting simulation and emulation environments to provide the network. A distributed controller environment is created with an Infinispan~\cite{marchioni2012infinispan} cache over a distributed network cluster. An elastic in-memory cluster architecture proposed in our previous work has been extended to provide a distributed adaptive execution of the controller~\cite{kathiravelu2014adaptive}.

Network flow routing algorithms that are commonly used in data center networks, such as the shortest path algorithm, are implemented as the base algorithms. SMART Algorithmic improvements were then applied on top of these base algorithms. As OpenDaylight follows the OSGi (Open Service Gateway Initiative)~\cite{gu2004toward} specification and offers a componentized modular architecture deployed on top of Apache Karaf, the controller extensions are developed as independent OSGi bundles and deployed alongside with the controller core bundles. The Model-Driven Software Engineering (MDSE) principles offered by the model-driven service abstraction layer (MD-SAL)~\cite{medved2014opendaylight} of OpenDaylight Lithium was leveraged in integrating the controller extensions and middlebox controllers. Algorithmic enhancements and extensions are deployed similarly. Due to the loose coupling in the design, SMART can be made to work with other controllers with minimal changes.
	\vspace{-2.5em}

\section{Preliminary Evaluation}
\label{sec:Eval}
SMART prototype was evaluated in a distributed simulation and emulation environment, on a cluster with 6 identical nodes (Intel\textregistered\ Core\texttrademark\ i7-2600K CPU @ 3.40GHz processor and 12 GB memory). Prototype implementation of SMART enhancements was compared with  base algorithms commonly used, to assess SLA fulfillment regarding priority flows, by extending the $xSDN$ software-defined networking enabled platform for network flow simulations~\cite{xsdn}. Experiments were carried out on multiple routing scenarios with the different SMART enhancement approaches, to evaluate the QoS, efficiency, and potential time and bandwidth overheads of SMART. Different type and number of flows with multiple different policies and intents were evaluated.
	\vspace{-1.1em}

\subsection{Long Running Flows}
The network was modeled as a small-world data center (SWDC) topology~\cite{shin2011small} with 1024 nodes and shortest path as the base routing algorithm, and flows were routed between chosen origin and destination nodes. Properties of the links, nodes, and flows were uniformly distributed, and congestion uniformly randomized. Network congestion was modeled by  dynamically making certain links slower to route. The routing process was repeated with SMART enhancements applied over the base routing algorithm, where the subflows of the priority flows were diverted. SLA was defined as the maximum time to complete the transfer of the priority flows. Slow links were marked from the descriptors of the simulation environment and failures and congestions were randomized across the links.

\begin{figure}[ht]
	\begin{center}
		\resizebox{0.9\columnwidth}{!}{
			\includegraphics[width=0.9\textwidth]{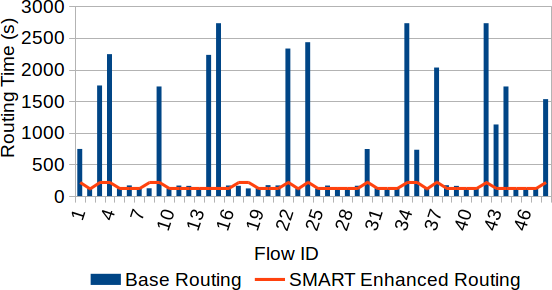}
		}
				\vspace{-1.8em}

	\end{center}
	
	\caption{SMART Divert Enhancement with Shortest-Path}
			\vspace{-1em}

	\label{fig:df}
\end{figure}

SMART was initially evaluated for  priority flows of longer duration. In a control experiment with no congestion, the flows took up to around 2 minutes (120 secs.) to complete, which is used as the approximate value for the soft-threshold in the experiment. The SLA limit was indicated as 250 secs. Figure~\ref{fig:df} shows the time taken for the routing with, and without the SMART enhancements for individual priority flows of equal lengths. This, across different origin and destination in multiple routing paths with or without congestion in certain links across the paths occurring randomly.

The complete time taken for the flow to reach the destination from origin is measured as the routing time. Though routing time is increased compared to the base routing algorithm without congestion, results show that SMART was able to avoid SLA violations in the data center, where the base algorithm violated the SLA for around 33\% of the priority flows in the presence of congestion in a few links. SLA violations were minimized for the selected priority flows by dynamically avoiding the slow routes as they are monitored and reported to the controller, based on previous packets  of the flow. Moreover, limited additional bandwidth consumption and controller CPU and memory load were observed. 
			\vspace{-0.8em}

In the regular routing, when a slow or congested route is encountered, the flow path cannot be changed dynamically, and the rest of the flow continues its routing in the original path regardless of a potential better alternative. In the SMART approach, after the specified time limit, SMART enhancer diverts or clones the subflow following the break point in an alternative route, which enhances the chosen base algorithm. While SMART depends on the availability of alternative routes in the network between the chosen node pairs, highly connected networks such as a mesh network typically offer a higher potential of finding an alternate route, that is as good as the shortest path, and should thus be leveraged.

\vspace{-0.5em}

\paragraph*{\textbf{Overheads of Flow Cloning}}

\begin{figure}[H]
	\begin{center}
		\resizebox{0.8\columnwidth}{!}{
			\includegraphics[width=0.8\textwidth]{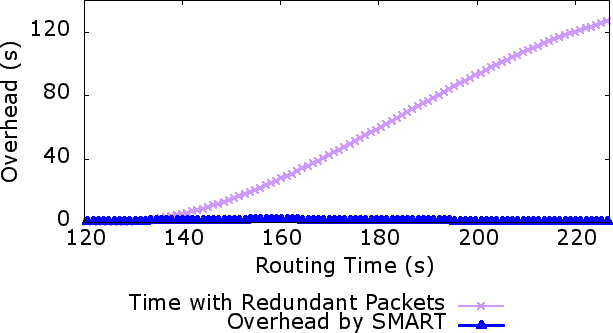}
		}
				\vspace{-2em}

	\end{center}
	\caption{Redundancy in Priority Flows}

	\label{fig:red}
\end{figure}

Similar routing times were observed when the experiments were repeated with cloning the subflows following the break point, instead of just  diverting. Redundancy in packets was monitored with flow cloning, and overhead imposed by SMART was measured them. Figure~\ref{fig:red} shows the routing time of the priority flows and time of routing with redundant packets due to the cloned subflows. It also shows the estimated overhead imposed by SMART on the routing.

				\vspace{-0.6em}

				\vspace{-1em}

				\vspace{-1.3em}

There were  no redundant packets till the soft limit is met and following that, the cloned packets assume an alternative route while the original flow continues. Hence a redundancy of up to 50\% of the entire routing time till the flow is recomposed, was observed, as shown by Figure~\ref{fig:red}. Further, Figure~\ref{fig:red} also indicates the overhead in routing time caused by SMART atop the base routing time without congestion. The overheads imposed by the SMART enhancements on the controller and switches are minimal for large flows, around 0.2\% of the routing time, which is often relatively a constant and negligible compared to the enhancements offered by SMART beyond 100\% in a congested route. 

\vspace{-1em}
\subsection{Short Running Flows}
\begin{figure}[ht]
	\begin{center}
		\resizebox{0.9\columnwidth}{!}{
			\includegraphics[width=0.9\textwidth]{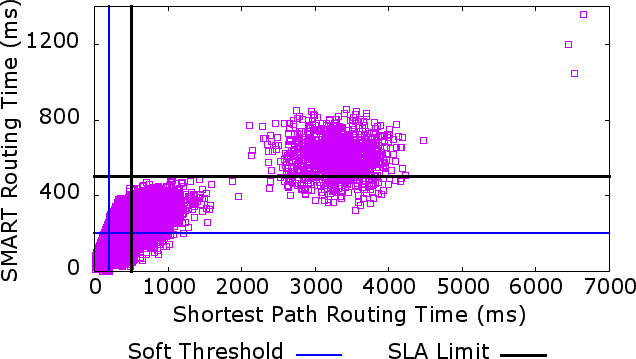}
		}
				\vspace{-1.7em}

	\end{center}
	\caption{SMART Clone Enhancements with Shortest-Path}

	\label{fig:short}
					\vspace{-0.5em}

\end{figure}

The experiment was repeated in the same data center simulation, modeled with up to 100,000 of short flows each consuming less than 1 second to complete its routing. Figure~\ref{fig:short} shows the time taken for a flow to route using shortest path as the base routing algorithm, as well as with SMART enhancements to clone the flows that exceed the soft-threshold. Thick and solid filled blocks in the diagrams indicate clustered outcomes for the pairs of base routing time vs SMART enhanced routing times, where thin and white blocks indicate single or less repeating pairs of observed values. During the congestion, an immediate overhead of around hundred milliseconds caused by SMART was observed. Yet SMART offered a speed up of up to 500\% in the presence of congestion. 

				\vspace{-1em}

\begin{figure}[ht]
	\begin{center}
		\resizebox{0.9\columnwidth}{!}{
			\includegraphics[width=0.9\textwidth]{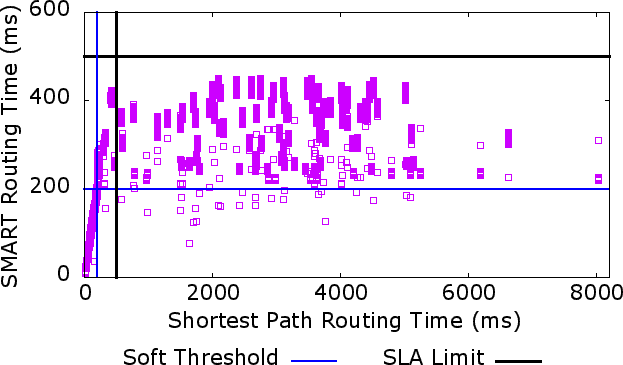}
		}
				\vspace{-1.7em}

	\end{center}
	\caption{SMART Clone and Replicate with Shortest-Path}

	\label{fig:repl}
					\vspace{-1.4em}

\end{figure}

SMART was configured to replicate the following flows at the origin when a flow of the same path reported a violation and cloned. Figure~\ref{fig:repl} indicates the time taken for SMART configured with this adaptive behavior, indicating there was no SLA violation with SMART enhancements.
					\vspace{-0.5em}

\begin{figure}[ht]
	\begin{center}
		\resizebox{0.9\columnwidth}{!}{
			\includegraphics[width=0.9\textwidth]{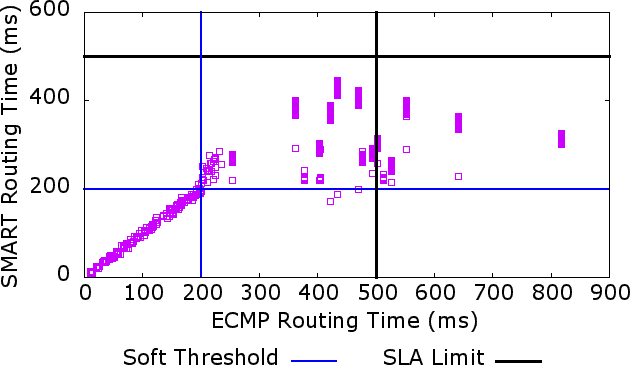}
		}
				\vspace{-1.7em}

	\end{center}
	\caption{SMART Clone and Replicate Enhancements with ECMP}

	\label{fig:all}
					\vspace{-0.8em}

\end{figure}

This experiment was repeated with equal-cost multi-path (ECMP) routing algorithm. Figure~\ref{fig:all} shows the time taken to route the flows in ECMP as the base algorithm in a congested network, with and without SMART enhancements.  As a base algorithm, ECMP distributes the flows across the alternatives. However, it is not aware of the congestion. Hence, SMART was able to enhance its performance by cloning the priority flows in an alternative route, which was readily available in ECMP, further replicating the following flows of the same path, that originally was found to be congested in an alternative route.

Considering all the cases, SLA violations were avoided by SMART by up to 95\%. The majority of the flows that originally violate SLA, abide to the SLA with the SMART enhancements. Performance of the controller and switches in detecting the violations, and updating the rules, contributes to the potential SLA violations. However, there is no flow which has an SLA violation with SMART enhancement, which is not also violated with the base routing. Unless the soft threshold was met, SMART enhancements were not invoked, as it indicated that the existing route was good enough to meet SLA and no potential congestion was foreseen.

				\vspace{-1em}

\paragraph*{\textbf{Assessment of Overheads}}

Bandwidth overhead was shown to be depending on multiple factors, such as the topology and size of the data center network, average length between any node pair, length of the congestion in the route (how many nodes and links are congested), location of the congestion and break point in the route and the flow, and the number of alternative routes available between any two nodes. 

SMART exhibits an adaptive behavior to the nature of the congestion, finding the right time to clone or divert. The contribution to congestion from cloning the subflows is minimal, as only around 16.7\% of the packets of the higher priority flows were found to be cloned in the typical data center network modeled, and hence the overall redundancy will be further smaller, depending on the fraction of the flows that are marked as higher priority to be cloned.

The logically centralized controller in an enterprise data center is a physically distributed cluster of high performance servers. Hence, the controller computations, such as determining the break point node and packet, monitoring the network flows for thresholds, imposing/changing the flow tables and policies in the relevant switches, and enforcing the SLA for the priority flows in the congested network based on the tags, are executed in the scale of microseconds. The overhead was estimated to be lower than a hundred milliseconds in switches when the break points are manipulated and flow tables are updated, with a minimal overhead in the bandwidth. As the base routing and FlowTagger are integrated with the SDN architecture, no overhead was caused by the deployment of SMART.

Mininet emulations of an about 1000-node data center with a distributed controller deployment of OpenDaylight over 6 nodes and SMART enhancements showed that the controller can handle the routing, rerouting, and reconstruction of flows and subflows effectively. This, without creating a bottleneck, as the majority of the decisions are handled by the nodes themselves with minimal intervention from the controller, unless a violation is triggered. Subflows still respect the ordering of packets. Hence, reconstruction of the original flow at the destination is straightforward, dropping the duplicate packets. The enhancements are adaptive to minimize the overhead even for much smaller flows, where if the performance improvement is minimal by cloning subflows, entire following/downstream  flows of the same priority, in the same path, will be replicated and routed in an alternative route along with the original route, or just rerouted in an alternative route omitting the slow route. 
				\vspace{-1em}

\section{Related Work}
\label{sec:related_work}
	\vspace{-1em}



While hosts in the data center networks are connected through multiple paths, TCP limits the connection to a single path. Multipath TCP (MPTCP) is a transport protocol that uses the available multiple paths between the nodes concurrently to route the flows across the nodes. MPTCP is proposed and implemented as an enhancement to TCP to improve the performance, bandwidth utilization, and congestion control through a distributed load balancing~\cite{raiciu2011improving}. MPTCP uses subflows in routing the flows, leveraging the multiple paths between the nodes in a network, and reconstruct the data in the destination in the original order~\cite{ford2013tcp}. 

Conga offers congestion-aware load balancing for data center networks through flowlet switching~\cite{alizadeh2014conga}. Flowlets are defined as the bursts or chunks of packets of a flow, that is separated with the other bursts of chunks by a gap~\cite{kandula2007dynamic}. Flows are often composed of flowlets and gaps between the flowlets, enabling an efficient partitioning of flows as flowlets and routing them in multiple alternative routes. While dynamically rerouting the network flows to optimize the bandwidth consumption has been proposed in the previous work~\cite{al2010hedera}, further research is necessary to enhance the existing networks and flow scheduling by leveraging the availability of the entire view in the central controller consisting of large computational power. 

Though data center networks are efficiently orchestrated and scaled with SDN, SLAs cannot be promised without dedicated and replicated resources. The existing work that leverages MPTCP or flowlets do not use redundant subflows for a reliable transfer of flows, or prioritize the flows based on user preferences to satisfy SLAs. Resending or cloning the flowlets if a previous flowlet has not reached the destination within the stipulated time should be researched. SLA-aware data center networks should be designed by exploiting the functionality offered by the middleboxes with minimally replicated resources and redundancy to ensure timely delivery of priority content. 
		\vspace{-1.8em}

\section{Conclusion and Future Work}
\label{sec:conclusion}
			\vspace{-0.6em}

SMART is developed as a fully functional middlebox-based approach for software-defined data center networks, by diverting or cloning subflows of priority flows for a timely delivery in a network with congested links. Preliminary evaluations on simulation and emulation platforms showed the efficiency of SMART in offering SLA-awareness to data center networks. As FlowTags effectively enforces policies regardless of the presence of middleboxes that modify the flow headers in the network, \projectName deployment is orthogonal to the presence of middleboxes. An ongoing development effort implements SMART on a real data center network and evaluate against CONGA and congestion-aware data center networks.


				\vspace{-0.6em}


\balance

\bibliography{references}

\end{document}